\documentclass[a4paper,11pt]{article}
\usepackage{jcappub}
\usepackage{lineno}
\usepackage{orcidlink}
\usepackage{ulem}

\usepackage{float}
\usepackage{amsmath}
\usepackage{graphicx}
\usepackage{graphicx}
\usepackage{indentfirst} 
\usepackage{amsmath}	
\usepackage{amssymb}

\usepackage{xcolor}

\newcommand{\mpch}{h^{-1} {\rm Mpc}}

\newcommand{\tttpt}{{$3\times 2$pt}}
\newcommand{\nside}{N_\mathrm{side}}
\newcommand{\lmax}{l_\mathrm{max}}
\newcommand{\glass}{\texttt{GLASS}}
\newcommand{\flask}{\texttt{FLASK}}

\newcommand{\wigner}{\texttt{wigner}}
\newcommand{\treecorr}{\texttt{Treecorr}}
\newcommand{\onecovariance}{\texttt{OneCovariance}}
\newcommand{\emcee}{\texttt{EMCEE}}
\newcommand{\ccl}{\texttt{CCL}}
\newcommand{\camb}{\texttt{CAMB}}
\newcommand{\class}{\texttt{CLASS}}

\title{Validation of Fast Mocks Generation for CSST Photometric Survey}
\author[~\orcidlink{0009-0009-3825-3816}1,2]{Yiming Hu,}
\author[~\orcidlink{0000-0002-9359-7170}2,3,4]{Yu Yu}

\affiliation[1]{\textit{Zhiyuan College, Shanghai Jiao Tong University, Shanghai 200240, China}}
\affiliation[2]{\textit{Department of Astronomy, School of Physics and Astronomy, Shanghai Jiao Tong University, 800 Dongchuan Road, Shanghai 200240, China}}
\affiliation[3]{\textit{State Key Laboratory of Dark Matter Physics, School of Physics and Astronomy, Shanghai Jiao Tong University, 800 Dongchuan Road, Shanghai 200240, China}}
\affiliation[4]{\textit{Key Laboratory for Particle Astrophysics and Cosmology (MOE),
\& Shanghai Key Laboratory for Particle Physics and Cosmology, Shanghai Jiao Tong University, Shanghai, 200240, China}}

\emailAdd{yuyu22@sjtu.edu.cn}

\abstract{
Weak lensing has become a powerful tool for probing the matter distribution in the Universe and constraining cosmological parameters. 
This paper aims to explore the fast mock generation pipeline to obtain the covariance matrix of the \tttpt{} analysis for the upcoming China Space Station Telescope (CSST).
We adopt the \glass{} pipeline, which generates matter distribution with lognormal assumptions, to create full-sky galaxy mocks with certain two-point statistics. We also employ the Markov-Chain Monte Carlo simulation to test the accuracy of the covariance matrix from the mock-generated galaxy catalogue.
Our work validates the accuracy of the \tttpt{} statistics in both spherical harmonic space and real space. The critical scale below which the fractional error of correlation exceeds 1\% can decrease as the resolution parameter $\nside$ increases. After excluding certain scales, the covariance matrix from the mock-generated galaxy catalogue can constrain the cosmological parameters with 1 \textperthousand \  accuracy. 
This work demonstrates the potential of \glass{} for real-space cosmological measurements and highlights the importance of discarding appropriate scales.}

\begin{document}
\maketitle

\section{Introduction}
\label{sec:intro}

When light passes through the gravitational field around a massive object, its path will be deflected.
Meanwhile, when the large-scale structure along its path distorts the light of a galaxy, the shape of the galaxy we observe will be slightly distorted, and this process is called weak gravitational lensing. By statistically analyzing the correlations between the shapes of galaxies, we can indirectly infer the projected matter distribution along the line of sight \cite{Kaiser:1991qi}. This measurement, combined with our models based on the $\Lambda$CDM cosmology, can be used to constrain the cosmological parameters \cite{RN100}, such as the density parameter for cold dark matter, $\Omega_c$, and the linear matter fluctuation amplitude within $8\,\mpch$, $\sigma_8$.

Through the weak gravitational lensing effect, Stage-\uppercase\expandafter{\romannumeral 3} surveys have been able to measure these cosmological constants with unprecedented precision, such as  the Kilo
Degree Survey (KiDS; \cite{RN89}), the Hyper Suprime-Cam survey (HSC; \cite{RN49}) and the Dark Energy Survey (DES; \cite{RN51}). However, these observations from the late Universe have shown discrepancies in the results of $\sigma_8$ compared to observations from the early Universe, namely those from the measurement of the Cosmic Microwave Background \cite{RN91,RN48,RN92,RN93,RN94,RN95,RN90}.
Lots of efforts have been made to address the so-called ``$S8$'' tension \cite{RN101,RN102}.
The upcoming Stage-\uppercase\expandafter{\romannumeral 4} surveys will further constrain these parameters in a wider, deeper, and more precise manner, such as \textit{Euclid} \cite{RN52}, Vera C. Rubin Legacy Survey of Space and Time (LSST; \cite{RN112}), and China Space Station Telescope (CSST; \cite{RN82}).

The CSST is a 2-m space telescope that will be in the same orbit as China’s manned space station, and it will cover an area of 17,500 $\text{deg}^2$ over 10 years \cite{RN82,RN53}. Equipped with the smaller spatial resolution, the wider wavelength coverage, and the capability of conducting both photometric and spectroscopic surveys, CSST will execute multiple scientific missions, including precisely measuring cosmological parameters with weak gravitational lensing. 


With the improvement of the statistical precision, weak lensing has many systematics to face. One of the systematics is from covariance estimation in the analysis. 
For instance, the dependence of covariance on cosmological parameters is proved to be important in $w$CDM model \cite{RN87}.
However, the $N$-body simulations are often computationally expensive, which cannot meet the requirements for covariance matrix estimation.
Sampling covariance for far from the $\Lambda$CDM model is still impossible due to the inherent complexity and the limitations of computational tools \cite{RN88}.
If the number of mocks $N_\text{mock}$ was just a factor of few larger than the size of the data vector $N$, the $N \times N$ covariance matrix would have a $O(\frac{1}{N})$ degradation \cite{RN86,RN85}. 
Recent mock generation methods that combine statistics and realistic physics have succeeded in weak gravitational lensing galaxy surveys. They possess both high efficiency and reliable accuracy in spherical harmonic space, such as the Full-sky Lognormal Astro fields Simulation Kit (\flask{}; \cite{RN16}) and the Generator for Large Scale Structure (\glass{}; \cite{RN8}). \glass{} is a simulation tool that generates galaxy samples by transforming Gaussian random fields into lognormal random fields, with the desired matter angular power spectrum as input. By employing a ``forward'' sequence, instead of the ``backward'' sequence in \flask{} \cite{RN16}, \glass{} accurately computed the angular power spectrum of the Gaussian field within a band limit (detailed in Section \ref{sub:glasspower}).

In this work, we explore the simulation pipeline for the covariance matrix of the \tttpt{}  analysis for CSST. We adopts the \glass{} simulation pipeline. The accuracy of \glass{} simulations has been verified in terms of the angular power spectrum \cite{RN8}, while this work focuses on testing the accuracy of two-point statistics in real space. Compared to the angular power spectrum, the angular correlation function, as a two-point statistic in real space, is often prone to being affected by theoretical systematics. Still, it is closer to actual observations and is more straightforward when considering selection effects and masks.

This paper is organized as follows: In Section \ref{sec:method}, we briefly review the process of generating galaxy samples using the \glass{} pipeline and test the statistics in spherical harmonic space first. In Section \ref{sec:results}, we present the results of the \tttpt{} analysis in real space and compare the impact of different spherical resolutions on the mock quality and parameter estimation. Finally, we summarize the results in Section \ref{sec:conclusion}.

\section{Method}
\label{sec:method}

\subsection{GLASS Overview}
\label{sec:glass}

\begin{figure}[t]
\centering
\includegraphics[width=0.8\linewidth]{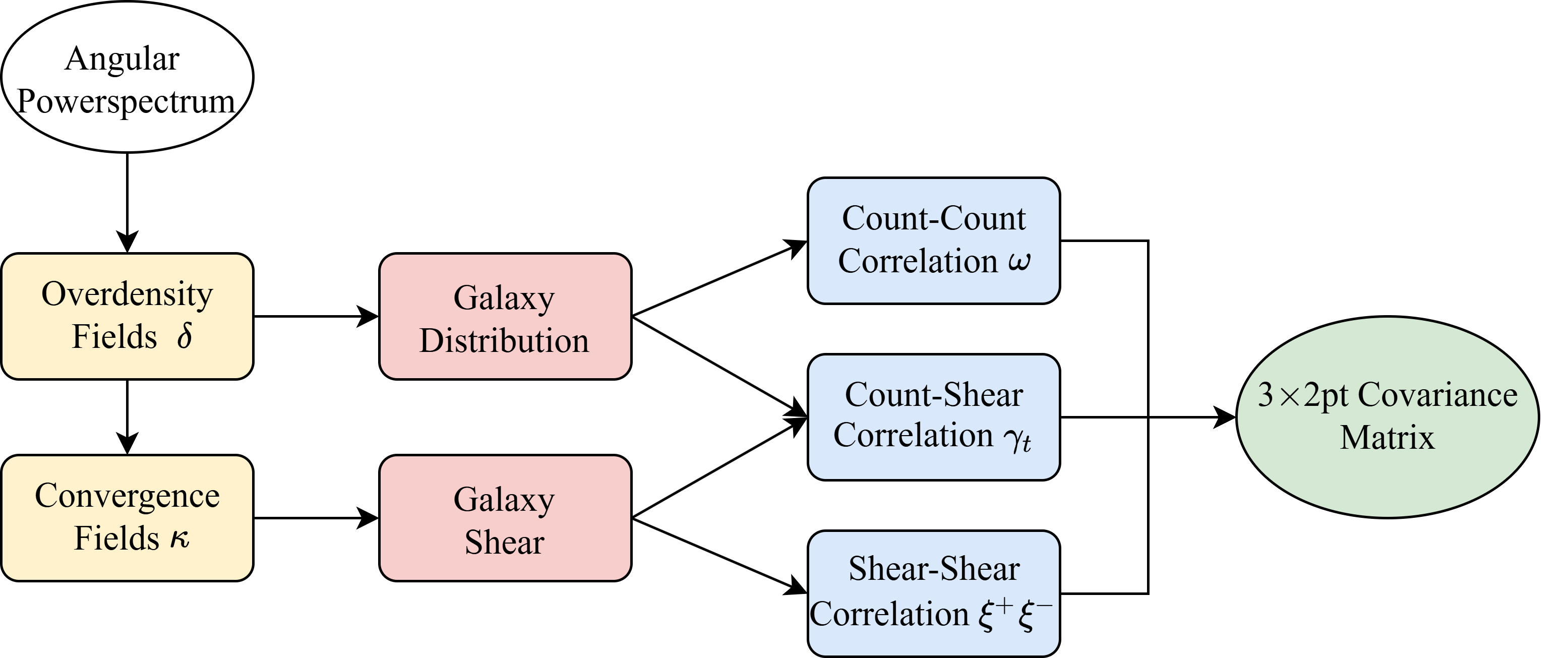}
\caption{\label{fig:flowchart}Flow chart of the pipeline adopted for generating our full-sky galaxy mocks and the \tttpt{} covariance matrix, based on the \glass{}.}
\end{figure}

In this work, we adopt the \glass{} pipeline largely consistent with the workflow in \glass{} \cite{RN8}, with some modifications to improve the performance and accuracy of the full-sky galaxy mocks generation. This pipeline creates lognormal fields from the input auto angular power spectra at different redshift bins. While the one-point statistics of the matter fields generated are lognormal and not an exact match to the real matter distribution \cite{RN106}, the approximation is widely adopted \cite{RN108,RN110,RN111}.  
Thus, the pipeline is capable of producing galaxy mocks with approximately correct one-point and two-point statistics.  
The outline of this workflow is shown in Figure~\ref{fig:flowchart}. In this section, we will briefly review how the \glass{} pipeline generates the galaxy catalogue with positions and shapes.

\subsubsection{Overdensity Fields}
\label{sec:glassdelta}

To generate a full sky galaxy catalogue, we first have to divide the simulated cosmic volume based on the redshift. We partition the universe into $n$ concentric spherical shells, named shell $i$ ($i = 1, 2, \cdots , n $), whose center is the observer. Starting with redshift $z_0 = 0$, the boundary of these shells are $z_0 < z_1 < \cdots < z_n$. Although the selection of the boundary $z_i$ is arbitrary, we will see in Section \ref{sec:gecorr} that the width of the redshift shells may affect the accuracy of the convergence fields. 
Striking an optimal balance regarding the width of the matter shell is crucial. On the one hand, overly wide shells can reduce the precision of the lensing map. On the other hand, excessively narrow shells tend to suppress the power spectrum on small scales\cite{RN133}. 

After fixing the boundary of each shell $i$, we can define the discretized overdensity field $\delta_i(\hat{u})$ for each shell $i$, which is the integral of the continuous overdensity field $\delta(z)$ along the line of sight $\hat{u}$,

\begin{equation}
\label{eq:delta}
\delta_i(\hat{u})=\frac{\int \delta\left(z,  \hat{u}\right) W_i(z) \mathrm{d} z}{\int W_i(z) \mathrm{d} z}\ ,
\end{equation}

\noindent where $W_i(z)$ is the matter weight function of shell $i$. Generally, there are several forms of $ W_i(z)$ to take. 
In this work, we choose the uniform weight in comoving distance,

\begin{equation}
W_i(z)= \begin{cases}1 / E(z) & \text { if } z_{i-1} \leq z<z_i\ , \\ 0 & \text { otherwise\ , }\end{cases}
\end{equation}

\noindent where $E(z)$ is the dimensionless Hubble function. 
With the matter weight function, we can then define the average redshift $\bar{z}_i$ of each shell $i$,

\begin{equation}
    \bar{z}_i = \frac{\int z W_i(z)dz}{\int W_i(z)dz}\ .
\end{equation}

For each shell $i$, we describe its two-point statistical properties through its auto-angular power spectrum $C_l$. In practice, the angular power spectrum can be calculated by various codes, such as \camb{} \cite{RN55,RN54}, \class{} \cite{RN56,RN57}, or \ccl{} \cite{RN58}. For our simulations, we only require the auto-correlation power spectra for each matter shell. Therefore, we opted to use \ccl{} for its efficiency and accuracy in computing the matter auto-correlation power spectra that we need. While the default \camb{} extension in \glass{} offers additional capabilities, our choice of \ccl{} is driven by the requirements of our current study.

In terms of one-point statistics, we assume that our discretized overdensity fields $\delta_i(\hat{u})$ are lognormal fields, which means that each field $\delta_i(\hat{u})$ is obtained by applying a lognormal transformation $f$ to a Gaussian random field $X_i$,

\begin{equation}
\label{eq:lognormal}
\delta_i(\hat{u}) = f(X_i(\hat{u}))=\lambda\left(\mathrm{e}^{X_i-\text{Var}(X_i)/2}-1\right)\ ,
\end{equation}

\noindent where $\lambda$ is the shift of the lognormal field, and $\text{Var}(X_i)$ is the variance of Gaussian field $X_i$. Considering the definition of overdensity, the overdensity $\delta_i = -1$ in the void region. Therefore, the value of shift parameter $\lambda$ is usually taken as $\lambda = 1$. Meanwhile, the $\text{Var}(X_i)/2$ term in Eq.~(\ref{eq:lognormal}) fixes the average value of overdensity $\bar{\delta}_i = 0$. 



\subsubsection{Power Spectrum of the Gaussian Fields}
\label{sub:glasspower}

As the spherical harmonic transform is nonlocal, the relation between the angular power spectrum of the Gaussian field $G_l$ and that of the lognormal field $C_l$ is non-trivial. However, the following relation exists between the angular correlation function of the Gaussian field $G(\theta)$ and that of the lognormal field $C(\theta)$, 

\begin{equation}
\label{eq:ctgt}
C(\theta)=\alpha \alpha^{\prime}\left[\mathrm{e}^{G(\theta)}-1\right]\ ,
\end{equation}

\begin{equation}
\label{eq:gtct}
G(\theta)=\ln \left[1+\frac{C(\theta)}{\alpha \alpha^{\prime}}\right]\ , 
\end{equation}

\noindent where parameter $\alpha = \left< X \right>+\lambda = \lambda$ and  $\alpha^{\prime} = \left< f(X) \right>+\lambda = \lambda$ since both the average value of the Gaussian field and that of the lognormal field are $0$.
We also have the relation between the angular power spectrum $C_l$ and the angular correlation function $C(\theta)$, 

\begin{equation}
\label{eq: clct}
C(\theta)=\sum_{l=0}^{\infty} \frac{2 l+1}{4 \pi} C_l P_l(\cos \theta)\ ,
\end{equation}

\begin{equation}
\label{eq: ctcl}
C_l=2 \pi \int_0^\pi C(\theta) P_l(\cos \theta) \sin (\theta) \mathrm{d} \theta\ ,
\end{equation}

\noindent where $P_l$ is the Legendre polynomial of degree $l$.

Combining Eq.~(\ref{eq:gtct}), Eq.~(\ref{eq: clct}), and Eq.~(\ref{eq: ctcl}), theoretically, we can obtain the angular power spectrum of the Gaussian field through the following ``backward'' sequence, 

\begin{equation}
C_l \rightarrow C(\theta) \rightarrow G(\theta) \rightarrow G_l\ .
\end{equation}

\noindent However, it is impossible to perform the summation of the infinite terms in Eq.~(\ref{eq: clct}) within the simulation. Once a band limit $\lmax$ on mode $l$ is imposed, the result of $G_l$ can be problematic. Therefore, in practice, the \glass{} pipeline employs the Gauss-Newton optimization algorithm with the ``forward'' sequence,

\begin{equation}
G_l \rightarrow G(\theta) \rightarrow C(\theta) \rightarrow C_l\ ,
\end{equation}

\noindent to ensure the accuracy of $G_l$ \cite{RN8}. 

\subsubsection{Sampling the Gaussian Fields on the Sphere}
\label{sec:glassgaussian}

A straightforward approach to sample a Gaussian random field $X_i$ on a sphere with specified two-point statistics is to sample its spherical harmonic coefficients $a_{lm}$ . This involves expanding the Gaussian field $X_i$ with the spherical harmonic functions $Y_{lm}(\theta, \phi)$,

\begin{equation}
X_i(\theta, \phi)=\sum_{l m} a_{l m} Y_{l m}(\theta, \phi)\ .
\end{equation}

\noindent Since we are sampling a real Gaussian random field, the coefficients $a_{lm}$ and their complex conjugates exhibit symmetry $a^*_{lm} = (-1)^m a_{lm}$, which means we only have to sample the coefficients $a_{lm}$ for $m \geq 0 $. 

It would be convenient if we sample the modulus $|a_{lm}|$ and phase $\varphi$ 
of the coefficients $a_{lm}$ separately,  $a_{lm} = |a_{lm}|\exp(i\varphi)\ $. The modulus $|a_{lm}|$ follows a Rayleigh distribution with scale parameter $\sigma$. According to the definition of the angular power spectrum, the scale parameter $\sigma$ of Rayleigh distribution is $\sigma = \sqrt{G_l/2}$. The phase follows a uniform distribution over $[0, 2\pi)$ for $m \neq 0$. For $m = 0$, the phase $\varphi$ follows a binomial distribution $\varphi = 0 \text{ or } \pi$ to ensure that $a_{l0}$ is real. 
In this work, we use the \texttt{Healpy} \cite{RN60,RN59} Python package to deal with the discrete spherical harmonic transformations on the discretized sphere.  


\subsubsection{Convergence Fields}
\label{sec:glasskappa}

In the case of weak lensing, it is usually assumed that the direction of light does not
change significantly during its propagation. Therefore, we adopt the Born approximation in this work to integrate the overdensity field $\delta(\hat{u} ; z)$ along the line of sight and build the convergence field $\kappa(\hat{u} ; z)$. Although ray-tracing is much closer to the real physical process than the Born approximation, recent research shows no significant difference between the two in terms of angular power spectra even from high-resolution simulations \cite{RN80}. Since our study focuses on two-point statistics instead of higher order statistics, the Born approximation remains a reasonable and efficient tool. The convergence field can then be written as 

\begin{equation}
\label{eq:kappa_0}
\kappa(\hat{u} ; z) 
 = \int_0^z \delta\left(\hat{u} ; z^{\prime}\right) W_L(z^{\prime}; z) \mathrm{d} z^{\prime}\ ,
\end{equation}

\noindent in which $W_L(z^{\prime};z)$ is the lensing kernel function, 

\begin{equation}
    W_L(z^{\prime};z) = \frac{3\Omega_m}{2}\frac{x_\text{M}(z)x_\text{M}(z^{\prime},z)}{x_\text{M}(z)}\frac{1+z^{\prime}}{E(z^{\prime})}\ .
\end{equation}

\noindent $x_{\mathrm{M}}$ is the dimensionless transverse comoving distance $x_{\mathrm{M}} = H_0d_{\mathrm{M}}/c$. $H_0$ denotes the Hubble parameter today, $c$ denotes the speed of light, and $d_{\mathrm{M}}$ denotes the comoving distance. 

Since we have already discretized the overdensity fields according to Eq.~(\ref{eq:delta}), the continuous integral Eq.~(\ref{eq:kappa_0}) must also be discretized, 


\begin{equation}
\label{eq:kappa_d}
\kappa_i(\hat{u})=\sum_{j=0}^{i-1} \delta_j(\hat{u})W_L(\bar{z}_j; \bar{z}_i) w_j\ ,
\end{equation}

\noindent where $w_i$ is the lensing weights which is an average over the matter weight function,

\begin{equation}
w_j=\frac{1}{W_j\left(\bar{z}_j\right)} \int W_j(z) \mathrm{d} z\ .
\end{equation}

In practice, to avoid redundant computations and reduce memory usage, the \glass{} pipeline adopted an iterative summation approach to calculate the convergence field \cite{RN104, RN103, RN105}. With this approach, we can only store the data of three adjacent fields in memory at the same time during the construction. Considering the massive memory consumption caused by a single field as the resolution $\nside$ increases, this approach effectively reduces the total memory required for the simulation.





\subsubsection{Shear Fields}
\label{sub:glassshear}

To describe the distortion of galaxy shapes due to weak lensing, the \glass{} pipeline employs a backward process of Kaiser-Squires inversion \cite{RN79} to generate the shear fields $\gamma(\hat{u})$. Firstly, we introduce the spin-weighted spherical harmonic functions $_sY_{lm}(\theta, \phi)$ and the spin-raising and spin-lowing operators $\eth$ and $\bar{\eth}$, 

\begin{equation}
\begin{aligned}
&\eth_s Y_{l m}=+\sqrt{(l-s)(l+s+1)}_{s+1} Y_{l m}\ ,\\
&\bar{\eth}_s Y_{l m}=-\sqrt{(l+s)(l-s+1)}_{s-1} \ Y_{l m}\ .
\end{aligned}
\label{eq:raise}
\end{equation}

\noindent Therefore, the Poisson equation between convergence field $\kappa(\hat{u})$ and the lensing potential $\psi(\hat{u})$ is 

\begin{equation}
\label{eq:kappa}
2 \kappa(\hat{u})= \eth \bar{\eth} \psi(\hat{u})\ .
\end{equation}

\noindent Meanwhile, by definition, the shear field $\gamma(\hat{u})$ is the spin-2 field of the lensing potential $\psi(\hat{u})$. Namely, the relation between shear field $\gamma(\hat{u})$ and the lensing potential $\psi(\hat{u})$ is

\begin{equation}
\label{eq:gamma}
2 \gamma(\hat{u})= \eth \eth \psi(\hat{u})\ .
\end{equation}

With Eq.~(\ref{eq:raise})-(\ref{eq:gamma}), we can establish the relationship between the spherical harmonic coefficients $\kappa_{lm}$ of the convergence $\kappa(\hat{u})$ and the spherical harmonic coefficients $\gamma_{lm}$ of the shear $\gamma(\hat{u})$, 

\begin{equation}
\gamma_{l m}=-\sqrt{\frac{(l+2)(l-1)}{l(l+1)}} \kappa_{l m} \quad\quad\quad (l\geq 2)\ .
\end{equation}

\noindent Since the shear $\gamma(\hat{u})$ is a spin-2 field, both the modes $l = 0$ and $l = 1$ vanish, $\gamma_{00} = \gamma_{1m} = 0$.

\subsubsection{Sampling the Galaxies} 
\label{sec:glassgalaxy}

In order to obtain the two-point correlation functions consistent with those in real surveys, we must generate realistic galaxy samples from the overdensity fields $\delta_{i}(\hat{u})$ and shear fields $\gamma_{i}(\hat{u})$. We adopt a commonly used linear galaxy bias model to generate our galaxies, 

\begin{equation}
    \delta^g_{i}(\hat{u}) = b_i\delta_i(\hat{u})\ ,
\end{equation}

\noindent where the number density contrast of galaxies $\delta^g_{i}(\hat{u})$ is proportional to the overdensity field $\delta_i(\hat{u})$ in the matter shell, with the proportionality coefficient being $b_i$. 
Overall, the relationship between $\delta^g$ and $\delta$ can be non-trivial due to the complicated physical processes of galaxy formation. 
However, a simple linear bias model here is sufficient to test the performance of the pipeline.

In practice, considering our overdensity fields are discrete grids on \texttt{Healpix} spheres, we sample the actual number of galaxies within each pixel by performing Poisson sampling, and we sample the specific positions of galaxies within each pixel through uniform sampling.
Following the sampling of galaxy positions, it is straightforward to characterize the lensing distortions in galaxy shapes. The observed distorted shapes are described by the reduced shear $g$, which is a combination of the convergence fields $\kappa(\hat{u})$ and shear fields $\gamma(\hat{u})$, 

\begin{equation}
    g = \frac{\gamma(\hat{u})}{1-\kappa(\hat{u})}\ .
\end{equation}

\noindent When the convergence field $\kappa \ll 1$, we have $g\approx \gamma$. Therefore, for each galaxy, the lensing distortion in its shape is the reduced shear $g$ within the pixel.

\subsection{Modifications and Implementation}
\label{sec:newclass}

In this work, we aim to generate a set of samples suitable for estimating the covariance matrix of CSST.
In this section, we will first introduce the basic information of our simulation, then explain the modifications we made to the pipeline, as well as the validation of the overdensity and convergence fields produced with these modifications. 

\subsubsection{Simulation Setup}
\label{sec:glassreview}

\begin{figure}[t]
    \centering
    \includegraphics[width = 1\linewidth]{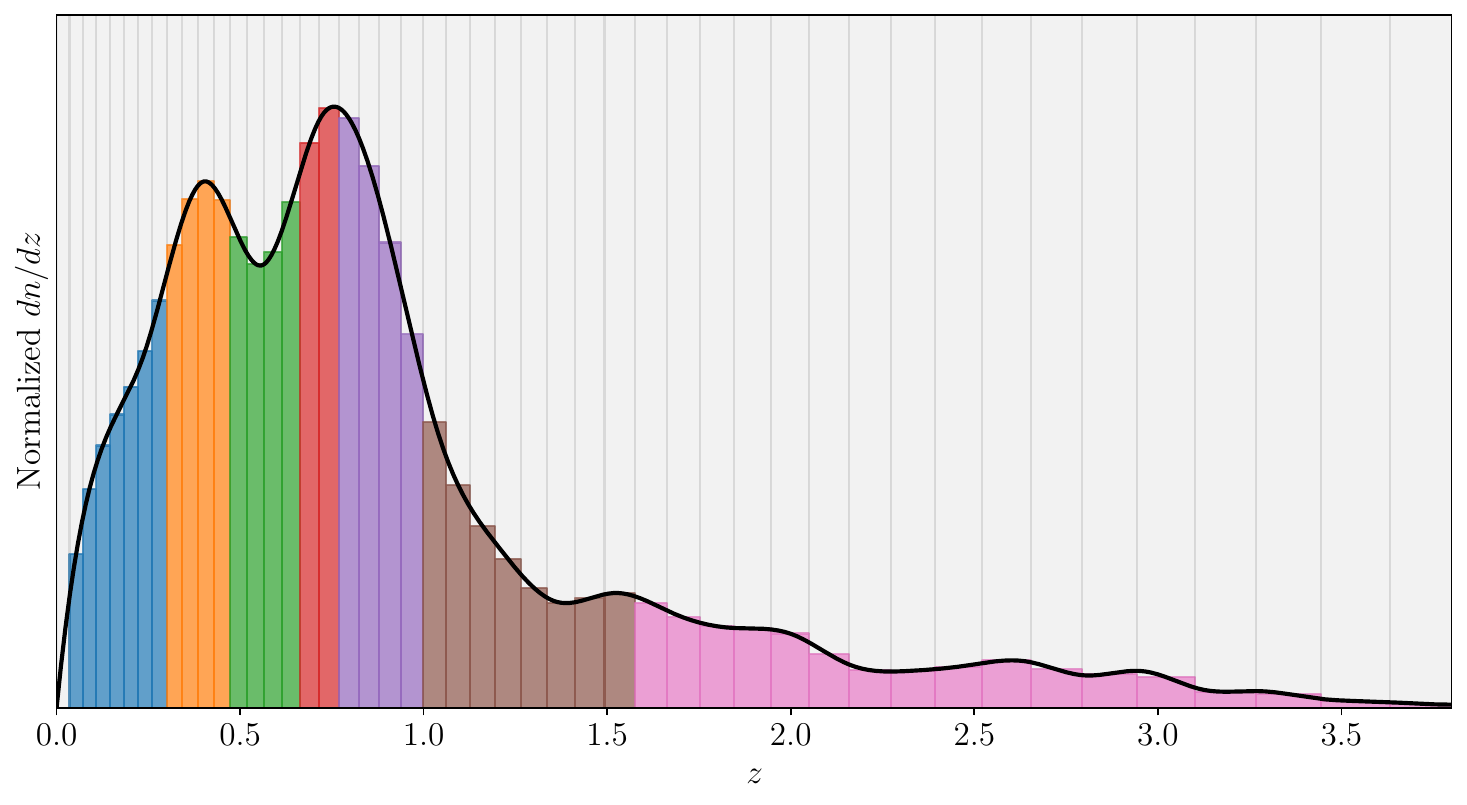}
    \caption{Matter shell boundaries (vertical grey lines) and the normalized galaxy redshift distribution (solid black line) adopted in this work. The colored areas represent the 7 tomographic bins, each containing approximately the same number of galaxies.}
    \label{fig:nz}
\end{figure}

In our simulations, we divide the redshift range $z \in (0, 3.7)$ into $48$ concentric matter shells based on equal comoving distances of $\Delta d_c = 150\text{ Mpc}$. Each matter shell is simulated using three different \texttt{HEALPix} spherical resolutions: $\nside = 1024\textit{, } 4096\text{, and } 8192$. The default settings in \texttt{Healpix} is $\lmax = 3\nside -1$. According to the results in \cite{RN8}, to generate reliable angular power spectrum $G_l$ through the ``backward'' sequence, it is suggested that the band limit should be between $\nside$ and $2\nside$. Therefore, we ultimately decided that the maximum spherical harmonic expansion multipoles for these resolutions are $2000$, $5000$, and $10000$, respectively.  
Within each matter shell, we adopt a constant bias parameter $b_i = 1$, and galaxies are distributed across the full sky according to the galaxy redshift distribution shown in Figure~\ref{fig:nz}. The redshift distribution here is the same as the one in \cite{RN68}, which is derived by applying the observational condition of the planned CSST survey to the galaxies with photo-z measurements from the COSMOS survey \cite{RN69,RN70}. For each resolution, 210 mocks are performed. As a general test, we use the following cosmological parameters as the fiducial model: $h = 0.7\text{, } \Omega_c = 0.25\textit{, }\Omega_b = 0.05\text{, and } \sigma_8 = 0.8$.

\subsubsection{Overdensity Fields}
\label{sec:newglassdelta}

One of the contradictions between mock construction and observations is that we want to directly measure the angular correlation functions in real space from the galaxy samples, yet the generation of spherical Gaussian fields is conveniently performed with the angular power spectrum in spherical harmonic space. Although the angular correlation function is merely a spherical harmonic transformation of the angular power spectrum, to achieve an angular resolution at the arcminute level, the maximum multipole of the angular power spectrum must reach at least $\lmax \sim 10^4$, which requires an enormous amount of computational cost. 

\begin{figure}[t]
    \centering
    \includegraphics[width = 1\linewidth]{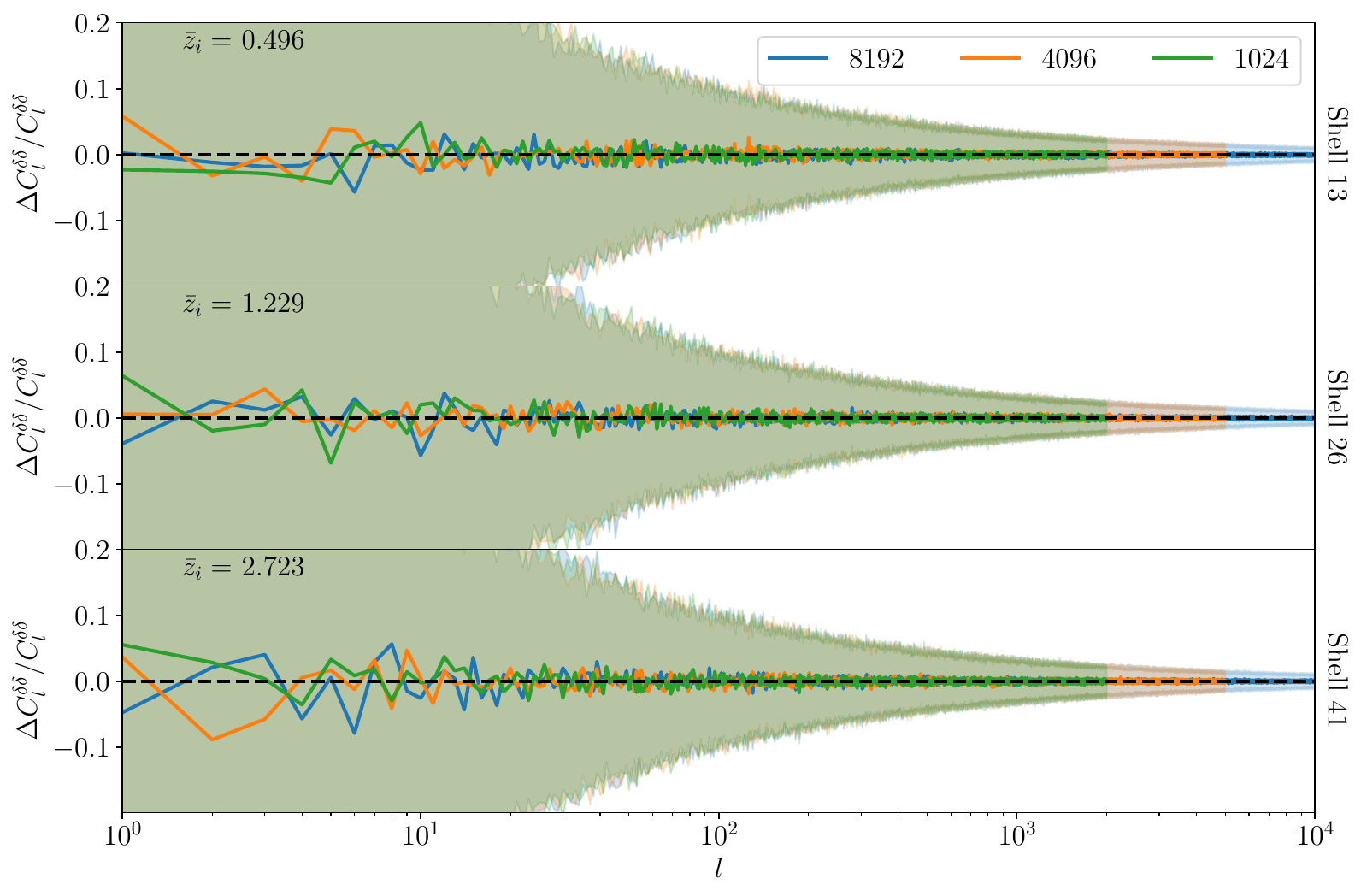}
    \caption{The fractional error of the angular power spectrum of the $\delta$ field in three shells. The colored lines represent the mean fractional error derived from 210 full-sky mocks, while the shaded area of each shows the 1-$\sigma$ interval. Due to the band limit, the fractional error is truncated at $l = 2000, 5000\text{, and }10000$. Within the band limit, there is good agreement between the angular power spectra of the generated overdensity fields and the ones of \ccl{} input. }
    \label{fig:overdensity}
\end{figure}

To test the two-point statistical properties of the overdensity fields generated by our pipeline after incorporating \ccl{}, we examined the auto-correlation angular power spectra of each overdensity field $\delta_i$. As shown in Figure~\ref{fig:overdensity}, the angular power spectra of the overdensity fields match the input \ccl{} angular power spectra very well at each spherical harmonic multipole, which is attributed to the accurate Gaussian angular power spectra calculated by the optimization algorithm of \glass{}.

\subsubsection{Convergence Fields}
\label{sec:newglasskappa}

Due to the definition of our discrete matter shells, the simulated convergence field is no longer an integral of a continuous overdensity field, but rather a superposition of discrete matter shells. The discretization inevitably leads to the systematics of the angular power spectrum of convergence fields. However, modifying the lensing kernel function can quantitatively mitigate these systematics.

According to Eq.~(\ref{eq:kappa_0}), the convergence field is essentially a superposition of the overdensity field. For the case of discrete matter shells, a discretized lensing kernel $\bar{W}_L(z;z^*)$ can then be defined as



\begin{equation}
    \bar{W}_L(z;z^*) \equiv W_L(\bar{z_i};z^*), \quad \text{ if } z_i \leq z < z_{i+1}\ .
\end{equation}

\begin{figure}[t]
    \centering
    \includegraphics[width = 1\linewidth]{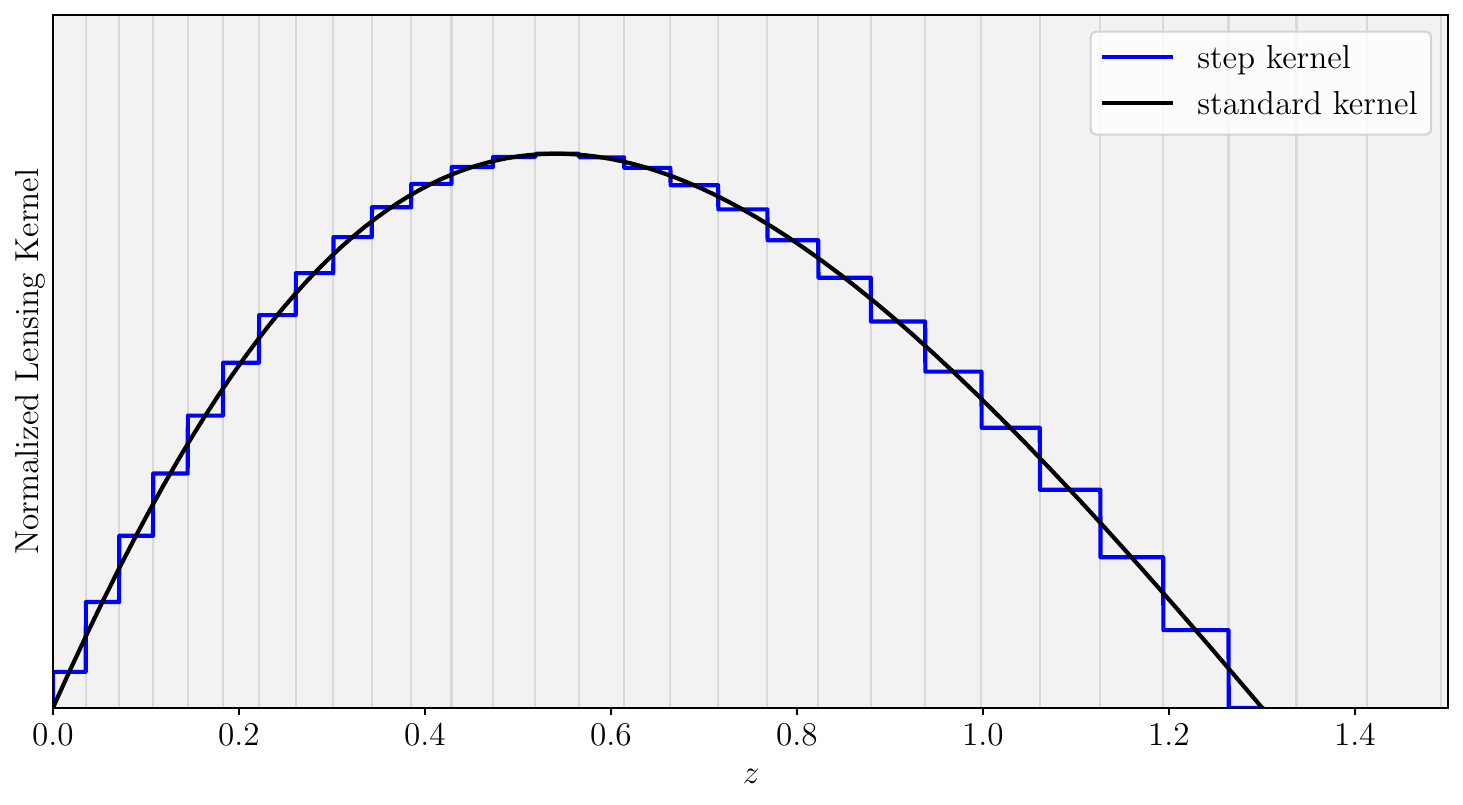}
    \caption{Lensing kernel of shell 27 ($z^*= 1.30$). The black line represents the standard lensing kernel $W(z;z^*)$ while the blue line represents the defined step lensing kernel $\bar{W}(z;z^*)$. Both $W(z;z^*)$ and $\bar{W}(z;z^*)$ have equal values at the average redshift $\bar{z}_i$ of each matter shell $i$.}
    \label{fig:kernel}
\end{figure}

\noindent Figure~\ref{fig:kernel} illustrates the difference between $W_L(z;z^*)$ and $\bar{W}_L(z;z^*)$ through a source at redshift $z^*=1.30$. Compared to $W_L(z;z^*)$, $\bar{W}_L(z;z^*)$ is a step function, and both $W_L(z;z^*)$ and $\bar{W}_L(z;z^*)$ have equal values at the average redshift $\bar{z}_i$ of each matter shell $i$.

\begin{figure}[t]
    \centering
    \includegraphics[width = 1\linewidth]{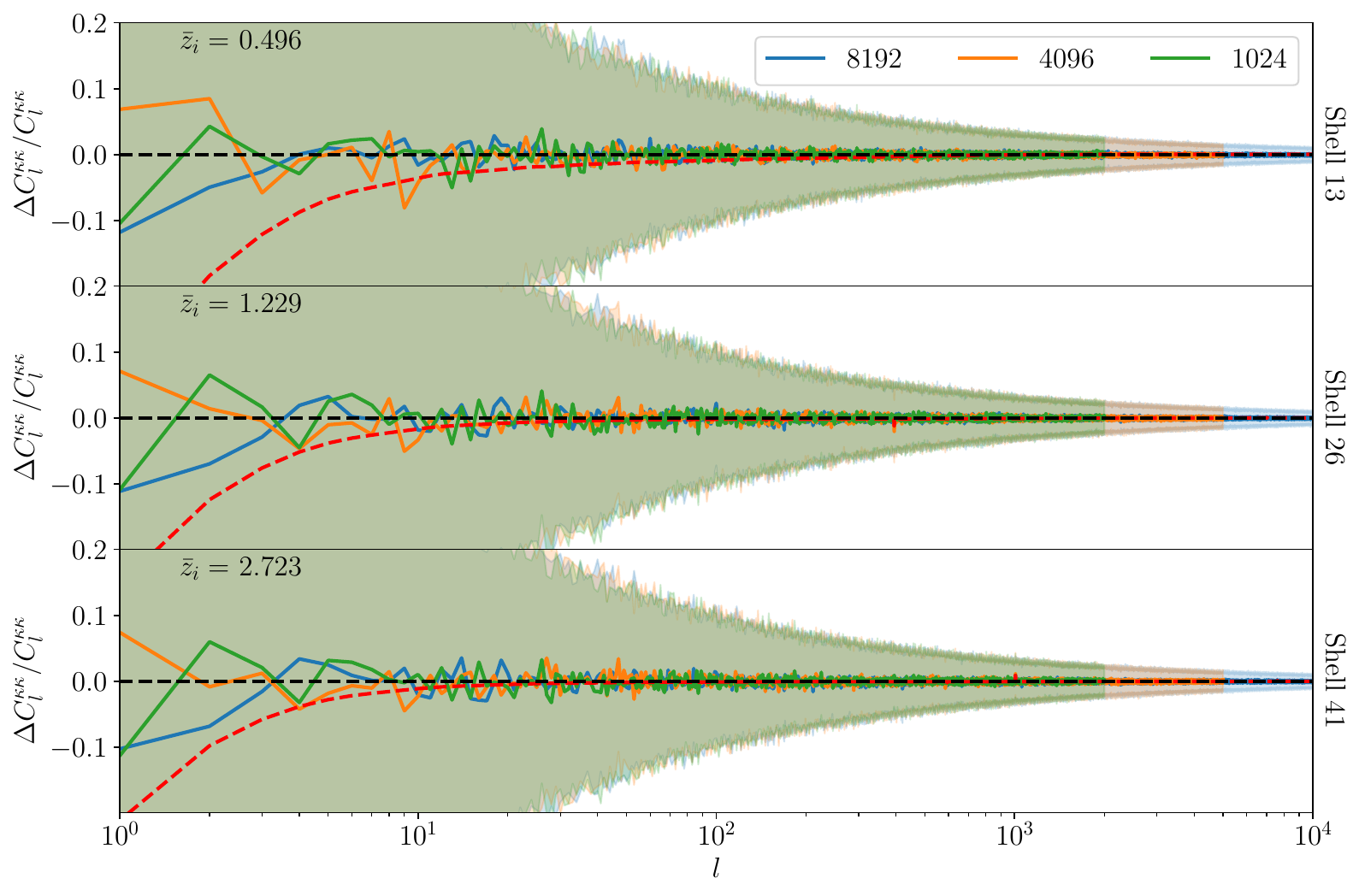}
    \caption{Same as Figure~\ref{fig:overdensity} but the solid lines show the result for lensing convergence field at three redshifts. The red dashed line represents the angular power spectrum difference predicted with the standard lensing kernel. The angular power spectra of generated convergence maps align well with theoretical values predicted by the discretized lensing kernel.
    }
    \label{fig:kl}
\end{figure}

Similar to the analysis of overdensity fields, we also measured the angular power spectrum of the convergence field and compared it with the theoretical power spectrum obtained by adopting discretized lensing kernel.
The result is shown in Figure~\ref{fig:kl}. By discretizing the lensing kernel, the angular power spectra of the generated field are in good agreement with the theoretical ones. 
On the contrary, using a continuous lensing kernel underestimates the lensing power spectrum, especially on the large scale. Meanwhile, as the redshift increases, the difference between the theoretical angular power spectra calculated by $W_L(z;z^*)$ and $\bar{W}_L(z;z^*)$ also gradually decreases.

One may feel nervous about the underestimation at the large scale of the theoretical angular power spectra calculated by $W_L(z;z^*)$ compared with the mean angular power spectra measured in the simulations in Figure~\ref{fig:kl}. In fact, the primary cause of this discrepancy lies in the lensing kernel at low redshifts. Notice that in Figure~\ref{fig:kernel}, the value of the step kernel at $z = 0$ does not start from zero and gradually increases; instead, it begins with a nonzero constant. This shape of lensing kernel is unphysical, yet it is naturally in agreement with Eq.~(\ref{eq:kappa_d}). This nonzero initial value can also be addressed by specifying a weight function that starts from zero, as done in the \glass{} pipeline \cite{RN8}. However, this requires an artificial selection of the weight function.
In this paper, we aim to minimize the introduction of artificial elements, and thus we have adopted the step lensing kernel with a nonzero initial value in the theoretical prediction.

\section{Results}
\label{sec:results}

To analyze the two-point statistics of the galaxy samples in real space, we employed a tomographic approach, dividing the galaxy samples from the $48$ matter shells into $7$ different redshift bins according to Figure~\ref{fig:nz}, maintaining approximately equal numbers of galaxies in each redshift bin.

In this section, we will first analyze the \tttpt{} for each tomographic bin, which includes the count-count correlation $\omega(\theta)$, the count-shear correlation $\gamma_t(\theta)$, and the shear-shear correlation $\xi^+(\theta)$ and $\xi^-(\theta)$. Then we will use the \tttpt{} between Bin 3 and Bin 5 as an example to sample the posterior distributions of $\Omega_m$ and $\sigma_8$ by running an MCMC analysis.

\subsection{Count-count Correlations}
\label{sec:ggcorr}

For the measurement of galaxy samples, we select the \treecorr{} \cite{RN25} Python package to compute the count-count correlation for the 210 mocks. We set the \treecorr{} parameters as follows: $\text{min\_sep} = 0.5 \text{ arcmin}$, $\text{max\_sep} = 300 \text{ arcmin}$, $\text{nbins} = 10$, and $\text{bin\_slop} = 0.01$. We also use the same parameters to calculate the count-shear correlation in Section \ref{sec:gecorr} and the shear-shear correlation in Section \ref{sec:eecorr}.
The following natural estimator is adopted to derive the count-count correlation,

\begin{equation}
    \xi = \frac{DD}{RR}-1,
\end{equation}

\noindent where $DD$ is the auto-correlation of the data and $RR$ is the auto-correlation of the random data. Here, we build a random catalogue which is approximately 10 times the size of the mock. 

\begin{figure}[t]
    \centering
    \includegraphics[width = 1\linewidth]{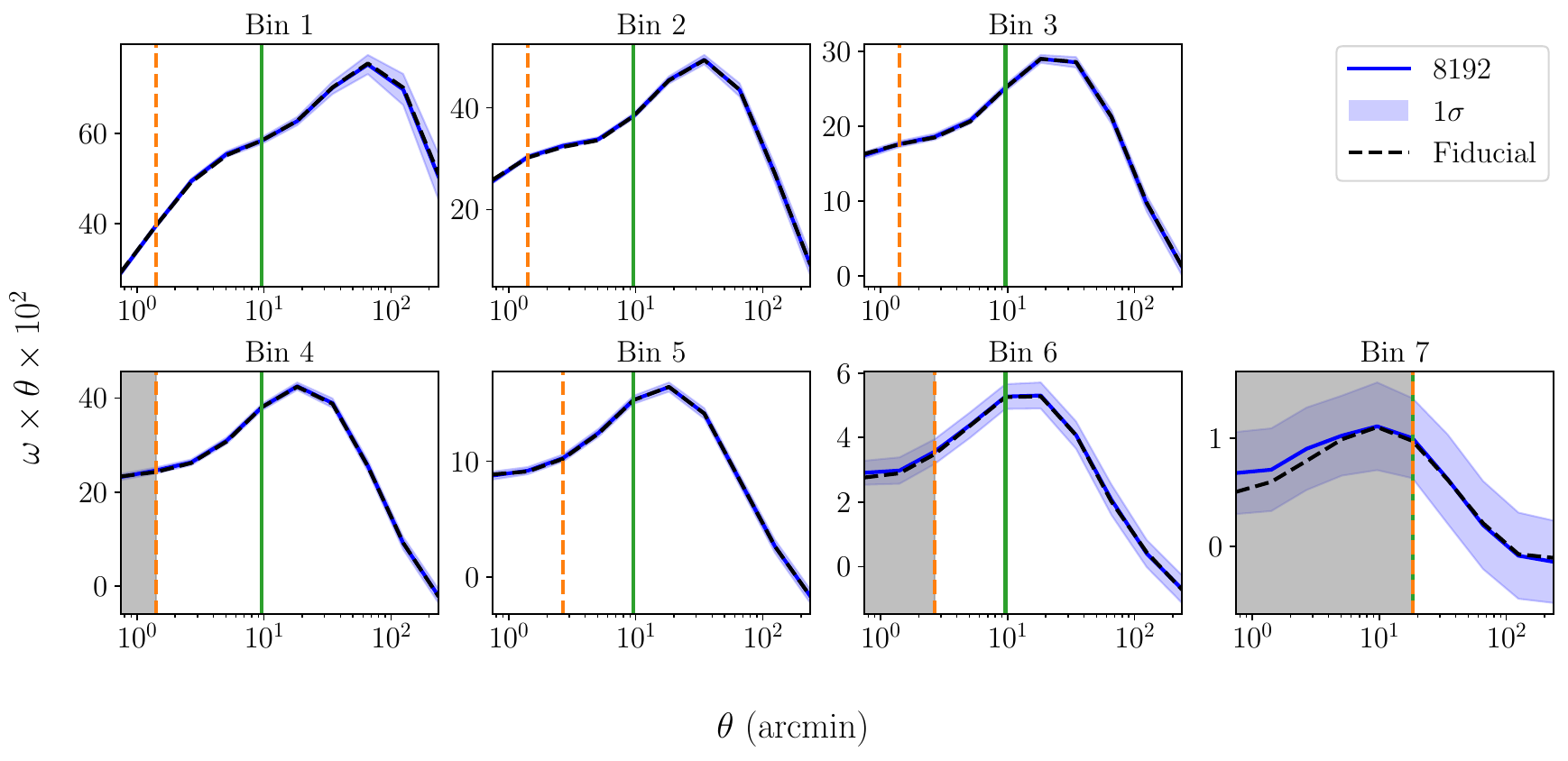}
    
    \caption{Count-count correlation functions of seven tomographic bins. In each panel, the blue solid line represents the mean value from 210 mocks with $\nside = 8192$, and the blue shaded area represents the 1-$\sigma$ interval. The black dashed line is the fiducial correlation function. The gray shaded regions indicate scales where the fractional error between the mean and theoretical values exceeds $1\%$ for $\nside = 8192$. The areas to the left of the orange dashed line and the green solid line represent the critical scales for $\nside = 4096$ and $1024$, respectively.}
    \label{fig:omega}
\end{figure}

To obtain the theoretical prediction of the count-count correlation, we first compute the theoretical angular power spectrum. According to Eq.~(\ref{eq: clct}), the galaxy count-count correlation $\omega(\theta)$ can be directly obtained by transforming the angular power spectrum $C_l$. However, in practice, the angular power spectrum must be cut off at a certain multipole $\lmax$, 
which will lead to a non-strict spherical harmonic transformation, thus causing systematics on small scales. Fortunately, the band limit $\lmax$ has already been introduced in the transformation of the angular power spectrum of a non-Gaussian field to a Gaussian one as mentioned in Section \ref{sub:glasspower}. 
Therefore, we can use the same $\lmax$ here, and thus the band limit will not compromise the consistency between the simulation and the prediction.
According to \cite{RN99}, the count-count correlation can then be written as 

\begin{equation}
\omega(\theta)=\sum_{l}^{\lmax} \frac{2 l+1}{4 \pi} C^{\delta\delta}_{l} P_{l}(\cos \theta)\ .
\end{equation}

\noindent Generally, computing Legendre polynomials of different orders involves high computational complexity. Here, we used the \wigner{} \cite{RN62} Python package to efficiently calculate Legendre polynomials. The \wigner{} package employs a recursive method to quickly obtain the values of Legendre polynomials and the Wigner d-matrix for various multipoles, which makes it ideal to calculate the transformation from angular powerspectrum to angular correlation.

In Figure~\ref{fig:omega}, we present the averaged count-count correlation functions of seven tomographic bins over 210 mocks with $\nside=8192$, along with the theoretical prediction.
The result shows that the mean value derived is substantially consistent with the theory.
However, we set a stringent requirement here.
The vertical grey bands indicate the region where the fractional error of the mean correlation function $\Delta\omega(\theta)/\omega(\theta)$ is larger than $1\%$.
For the mocks with $\nside = 1024$ and $ 4096$, the maximum scales where the fractional error exceeds 1\%, referred to as the critical scales, are also marked with green solid and orange dashed lines in Figure~\ref{fig:omega},  respectively.

For the same tomographic bin, the critical scale decreases as $\nside$ increases. This is because mocks with a larger $\nside$ also have a larger $\lmax$, so the spherical harmonic transform is more accurate on small scales. For the same $\nside$, different tomographic bins have different critical scales. For $\nside = 4096$, the critical scale increases from about $1 \ {\rm  arcmin}$ for Bin 1 to about $10 \ {\rm arcmin}$ for Bin 7. Similarly, for $\nside = 8192$ and $1024$, the critical scale also tends to increase with redshift. This is because the absolute value of the correlation function generally decreases as the redshift increases, causing the fractional error to increase with redshift.

\subsection{Count-shear Correlation}
\label{sec:gecorr}

Theoretically, the count-shear correlation is the cross-correlation between overdensity and shear. The overdensity has spin 0, while the shear field has spin 2. Therefore, the count-shear angular correlation function $\gamma_{t}(\theta)$ is related to the angular power spectrum $C_{l}^{\delta \kappa}$ through the following transformation \cite{RN99}, 

\begin{align}
\gamma_{t}(\theta) & =\sum_{l}^{\lmax} \frac{2 l+1}{4 \pi} C_{l}^{\delta \kappa} d^{l}_{2,0}(\theta) &\text{(full-sky)}&\ , \label{eq:3_3}\\
& =\int_0^{\lmax} \frac{d \ell \ell}{2 \pi} C_{\ell}^{\delta \kappa} J_2(\ell \theta) & \text{(flat-sky)}&\ , \label{eq:3_4}
\end{align}


\noindent where $d_{2,0}^{l}(\theta)$ is the Wigner (small) d-matrix $d_{m,m^{\prime}}^{l}$ with $m = 2$ and $m^{\prime} = 0$, and $J_2$ is the Bessel function of order 2. For similar reasons as discussed in the Section \ref{sec:ggcorr}, we imposed a band limit in Eq.~(\ref{eq:3_3}) and Eq.~(\ref{eq:3_4}). It should be noted that, in the simulation, the band limit for the count component arises from the transformation of the Gaussian power spectrum to a non-Gaussian power spectrum in Section \ref{sub:glasspower}, while the band limit for the shear component originates from the Kaiser-Squires inversion employed in Section \ref{sub:glassshear}. Therefore, the introduction of band limit in Eq.~(\ref{eq:3_3}) and Eq.~(\ref{eq:3_4}) will not lead to significant inconsistency between the simulation and the theory.
In most cases, such as in a small patch of sky, Eq.~(\ref{eq:3_4}) serves as a good approximation. However, in our simulations, we generated full-sky data and required our correlation functions to be accurate on 1\% scales. Therefore, despite the higher computational complexity of the Wigner d-matrix compared to the Bessel function, we still use Eq.~(\ref{eq:3_3}) to compute the count–shear correlation.

\begin{figure}[t]
    \centering
    \includegraphics[width = 1\linewidth]{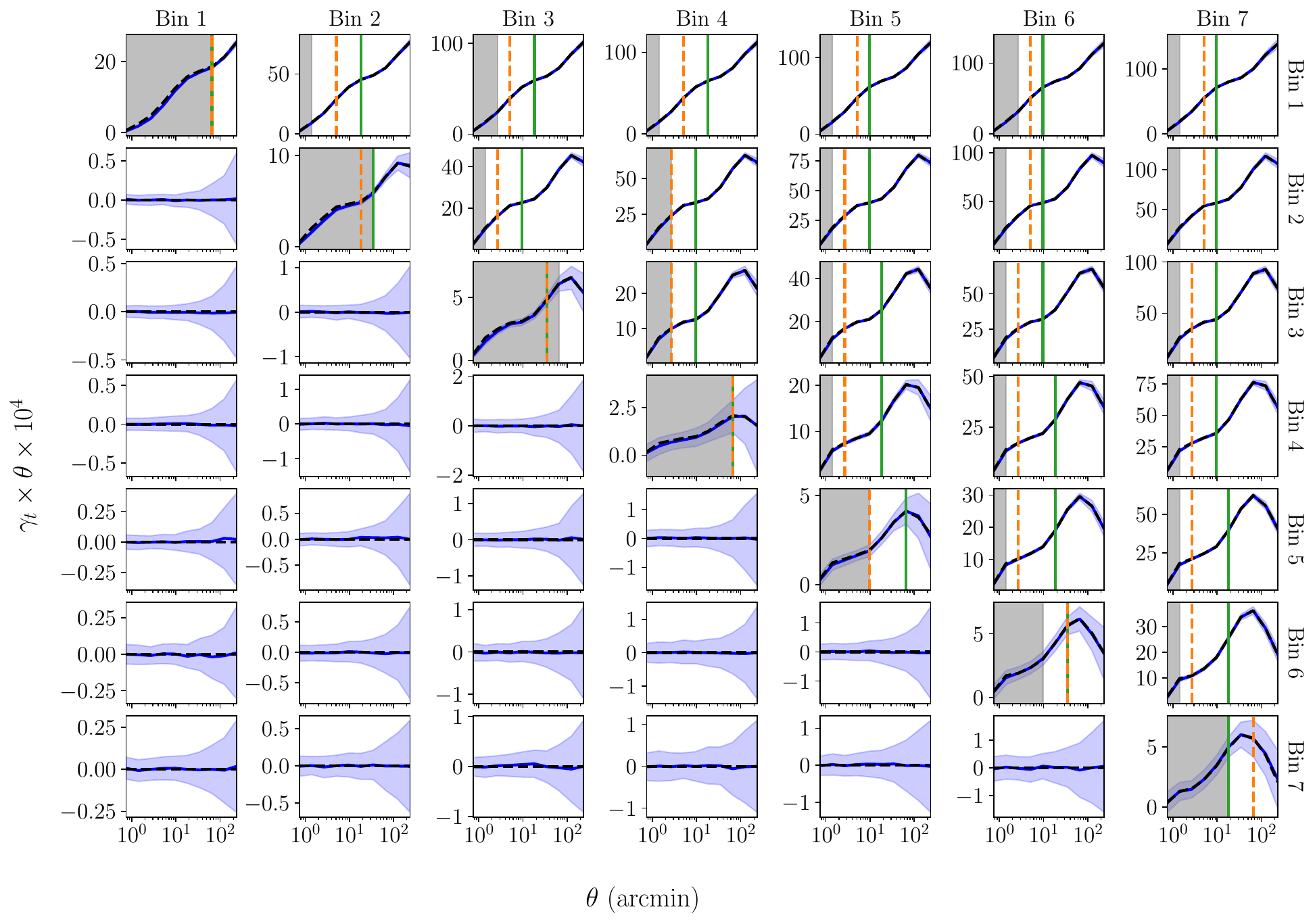}
    \caption{Same as Figure~\ref{fig:omega} but for the count-shear correlation functions of seven tomographic bins.}
    \label{fig:7}
\end{figure}

Figure~\ref{fig:7} shows the count-shear correlation between seven tomographic bins. The legends are the same as those in Figure~\ref{fig:omega}. The correlations in the panels located below the diagonal are all close to zero because the lens galaxies are behind the source galaxies, and thus, there is no physical correlation between them. 

For the panels where the lens galaxies are in front of the source galaxies, the mean value derived from 210 mocks at $\nside = 8192$ shows good consistency with the theoretical prediction. The main discrepancies occur in the auto- and cross-correlations of Bin 1 to Bin 3, where the measured values are lower than the theoretical values. This underestimation arises from the discretization of the matter shells. For the innermost several bins, the number of matter shells in front of them is too small to accurately correct for the errors introduced by the discrete redshift intervals. Adopting a non-uniform comoving distance for the division of matter shells, with more matter shells allocated at low redshifts, might help to mitigate this discrepancy. 

The critical scale is still affected by the resolution parameter $\nside$.
As shown in Figure~\ref{fig:7}, reducing $\nside$ continues to shift the critical scale toward larger scales.
Meanwhile, for the same $\nside$, the relationship between the critical scale of count-shear correlation and redshift is no longer monotonic. On the one hand, as redshift increases, the signal of the count-shear correlation decreases gradually, which thus causes the fractional error to increase. On the other hand, as redshift increases, the number of matter shells in front of the selected tomographic bin also increases, making the shear component more accurate. These two effects work together, leading to a non-monotonic relationship between the critical scale and redshift.

\subsection{Shear-shear Correlation}
\label{sec:eecorr}

Given that the shear-shear correlation is the correlation of two fields with spin 2, it is made of two different components, namely $\xi^+(\theta)$ and $\xi^-(\theta)$.
Therefore, we calculate the theoretical prediction of $\xi^+(\theta)$ and $\xi^-(\theta)$ through the following transformation \cite{RN99}, 

\begin{align}
\xi^{\pm}(\theta) & =\sum_{l}^{\lmax} \frac{2 l+1}{4 \pi} C_{l}^{\kappa \kappa} d^{l}_{2,\pm 2}(\theta) &\text{(full-sky)}&\ , \label{eq:3_5}\\
& =\int_0^{\lmax} \frac{d \ell \ell}{2 \pi} C_{\ell}^{\kappa \kappa} J_{2\mp 2}(\ell \theta) & \text{(flat-sky)}&\ , \label{eq:3_6}
\end{align}


\noindent where $d_{2,\pm 2}^{l}(\theta)$ is the Wigner (small) d-matrix $d_{m,m^{\prime}}^{l}$ with $m = 2$ and $m^{\prime} = \pm 2$, and $J_{2\mp2}$ is the Bessel function of order $0$ and $4$, respectively.
Consistent with Section \ref{sec:ggcorr} and Section \ref{sec:gecorr}, we have added the band limit in Eq.~(\ref{eq:3_5}) and Eq.~(\ref{eq:3_6}), and we use Eq.~(\ref{eq:3_5}) in the actual calculations.

\begin{figure}[t]
    \centering
    \includegraphics[width = 1\linewidth]{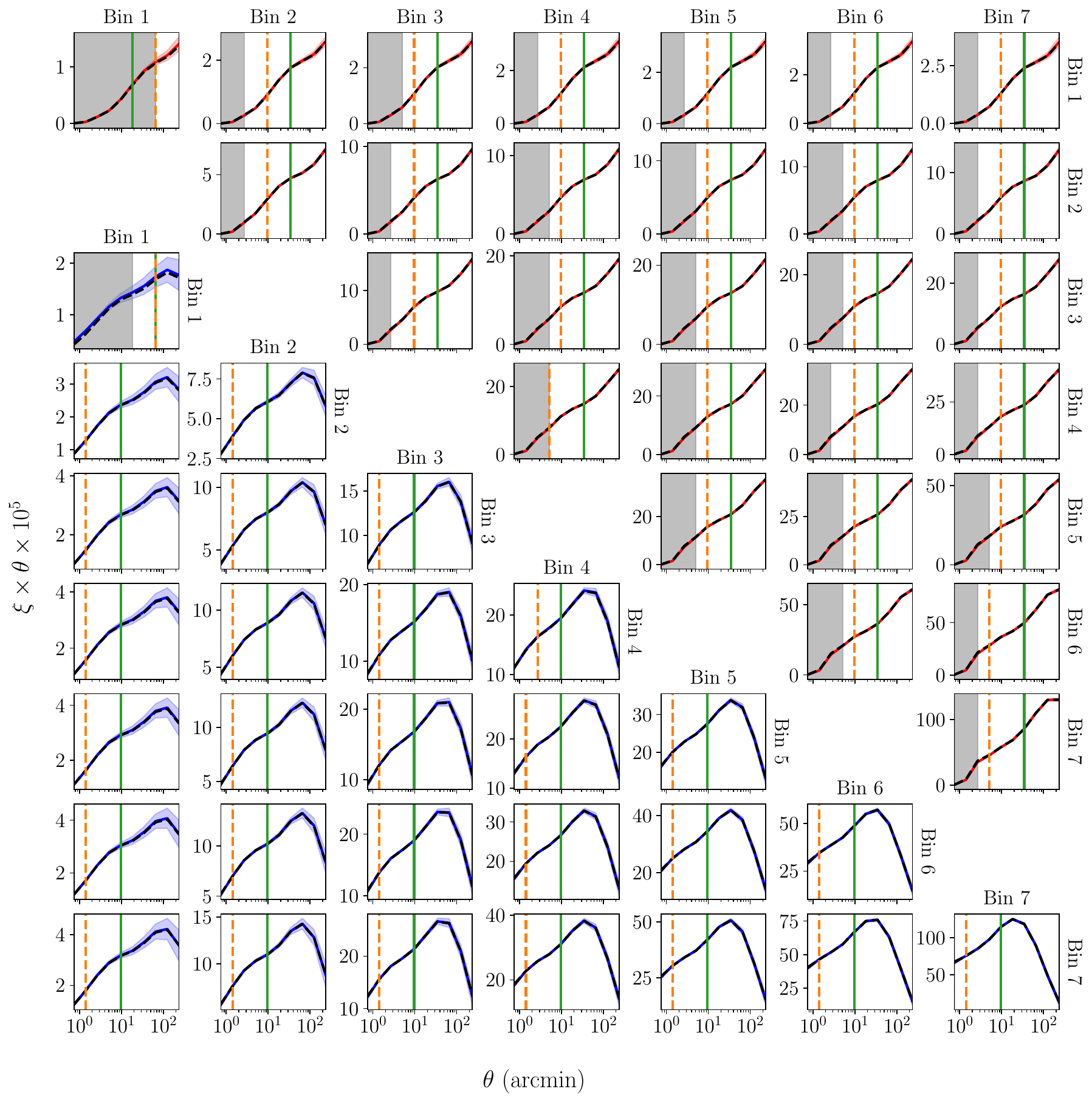}
    \caption{Same as Figure~\ref{fig:omega} but for the shear-shear correlation functions of seven tomographic bins. The lower left panels represent $\xi^+$(blue curves), and the upper right panels represent $\xi^-$(red curves)}.
    \label{fig:8}
\end{figure}

Figure \ref{fig:8} shows the shear-shear correlation between the seven tomographic bins. The lower left panels represent $\xi^+$ (blue curves), and the upper right panels represent $\xi^-$ (red curves). Our measurements of the mean values of both $\xi^+$ and $\xi^-$ show good agreement with theoretical values in magnitude and trend. 

Overall, we can find that the critical scales for $\xi^-$ are generally larger than those for $\xi^+$. This is due to the distinct oscillatory behavior of $d_{2,4}^l(\theta)$ and $d_{2,0}^l(\theta)$ in Eq.~\ref{eq:3_5} at small scales. The finite band limit $l_{\text{max}}$ causes the spherical harmonic transform to lose accuracy beyond a certain scale, while the stronger oscillatory nature of $d_{2,4}^l(\theta)$ amplifies this discrepancy, causing $\xi^-$ to deviate from the theoretical value at larger scales.

For a given $\nside$, except for the $\xi^+$ and $\xi^-$ of Bin 1 $\times$ Bin 1, the critical scales do not change significantly with redshift. Bin 1 is located in the innermost part of the entire simulation, and thus the correlation of Bin 1 primarily arises from its own lensing effect within the bin.
For other bins, the critical scale mainly depends on the band limit. A larger $\nside$, which means a larger band limit, can thus result in a smaller critical scale.

\subsection{MCMC Analysis}
\label{sec:mcmc}

To test the reliability of simulations with different resolution parameters in estimating the covariance matrix, we calculated the covariance matrices for 210 simulations at three different resolutions and attempted to constrain the cosmological parameters using the MCMC analysis. 
As a toy test, we have not introduced any potential observational effects throughout the process, like masks, inhomogeneity, photometric redshift, or intrinsic alignment, as well as the shape noise, which would lead to a very strong constraint on cosmological parameters within a small contour.

We choose the $\sigma_8$-$\Omega_m$ plane as the constraint space, since the weak lensing effect is quite sensitive to matter clustering. We employed the likelihood $\mathcal{L}(\mathbf{x})$ of the multivariate normal distribution,

\begin{equation}
    \mathcal{L}(\mathbf{x}) \propto \exp \left(-\frac{1}{2}(\mathbf{x}-\mathbf{\mu})^T\Sigma^{-1}(\mathbf{x}-\mathbf{\mu})\right).
\end{equation}

\noindent Generally, to ensure the covariance matrix $\Sigma$ is invertible, the number of independent mocks $N_{\text{mock}}$ should be much larger than the length of the data vector $N$.  
In our simulations, $N_{\text{mock}} = 210$. Limited by the finite number of simulations, 
we only analyzed the \tttpt{} between two selected tomographic bins, namely Bin 3 and Bin 5. Specifically, the data vector is sequentially composed of $\omega_{3\times3}$, $\gamma_{t3\times5}$, $\xi^+_{5\times5}$, and $\xi^-_{5\times5}$, which resulted in the length of the data vector being $40$. 

\begin{figure}[t]
    \centering
    \includegraphics[width = 1\linewidth]{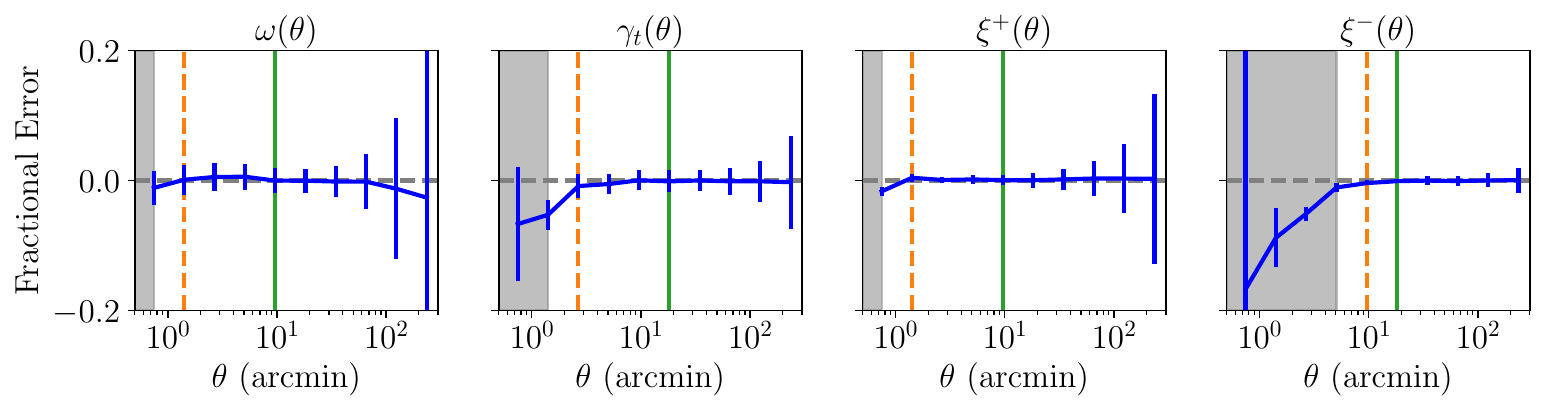}
    \caption{The fractional error of the data vector. From left to right, the four panels depict the fractional errors of $\omega(\theta)$ for $\text{Bin }3 \times \text{Bin }3$, $\gamma_t(\theta)$ for $\text{Bin }3 \times \text{Bin }5$, and $\xi^+(\theta)$ and $\xi^-(\theta)$  for $\text{Bin }5 \times \text{Bin }5$. Each correlation function represents the average at $\nside = 8192$ and the error bars represent 0.1 times the expected sigma for CSST. 
    The gray shaded regions, the regions to the left of the orange dashed line, and the regions to the left of the green solid line represent the discarded regions for $\nside =8192, 4096,$ and $1024$, respectively.}
    \label{fig:data vector}
\end{figure}

We use the mean correlation measured from the 210 mocks with $\nside=8192$ as the observation.
The underlying cosmology for this observation is $\Omega_m=\Omega_c+\Omega_b = 0.30$ and $\sigma_8 = 0.8$. 
Figure~\ref{fig:data vector} shows the fractional error of the mean value compared to the theoretical prediction. In the following process, we discard scales below the critical scales. In Figure~\ref{fig:data vector}, the gray shaded regions, the regions to the left of the orange dashed line, and the regions to the left of the green solid line represent the discarded regions for $\nside =8192$, $4096$ and $1024$, respectively.

\begin{figure}[t]
    \centering
    \includegraphics[width = 1\linewidth]{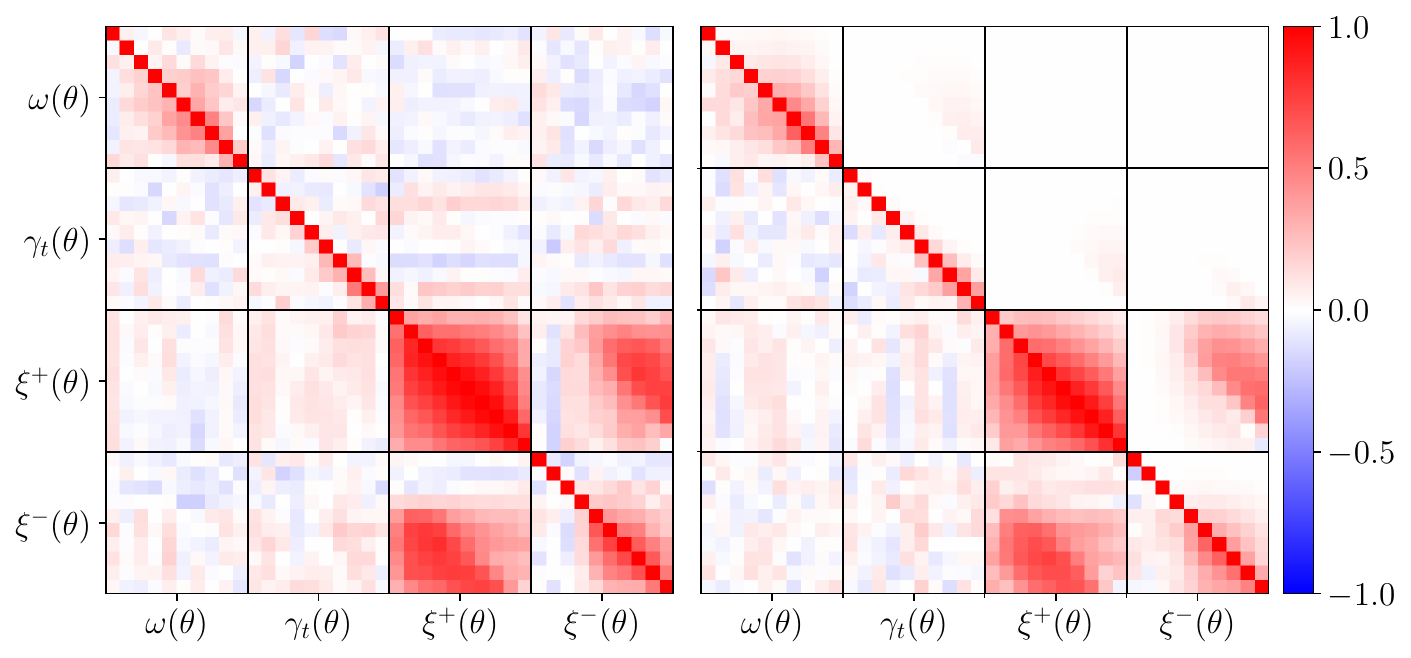}
    \caption{
The covariance matrix calculated by OneCovariance (upper right of the right panel) and the covariance matrix from 210 mocks at $\nside = 1024$ (lower left of the left panel), $4096$ (upper right of the left panel) and $8192$ (lower left of the right panel). The matrices are normalized by their diagonal elements for clarity.}
    \label{fig:covariance}
\end{figure}

As an additional verfication, we also used the \onecovariance{} \cite{RN63} Python Package to estimate the covariance matrix. \onecovariance{} is a tool for calculating the theoretical covariance matrix of large-scale structure surveys like KiDS, supporting the computation of various observables in both real and fourier spaces, especially the correlation functions of \tttpt{}.  Figure \ref{fig:covariance} displays the theoretical covariance matrix calculated by \onecovariance{} and the covariance matrix estimated from the $210$ simulations at $\nside = 1024$, $4096$ and $8192$. 
The number of mocks $N_{\text{mock}}$ is only about five times the length of the data vector $N$, so there is $O(1/N)$ Gaussian noise in the covariance matrix, which is more apparent for the parts that are near zero.
However, in general, the results are still consistent in overall structure, especially near the diagonal.

With the help of the \emcee{} \cite{RN64} Python Package, we performed MCMC sampling with the covariance matrices sampled at three different resolutions and the covariance matrix estimated by \onecovariance{}. We used $20$ walkers, each sampling approximately $5000$ steps to explore the plane of the two cosmological parameters $\sigma_8$ and $\Omega_m$. 

\begin{figure}[t]
    \centering
    \includegraphics[width = 1\linewidth]{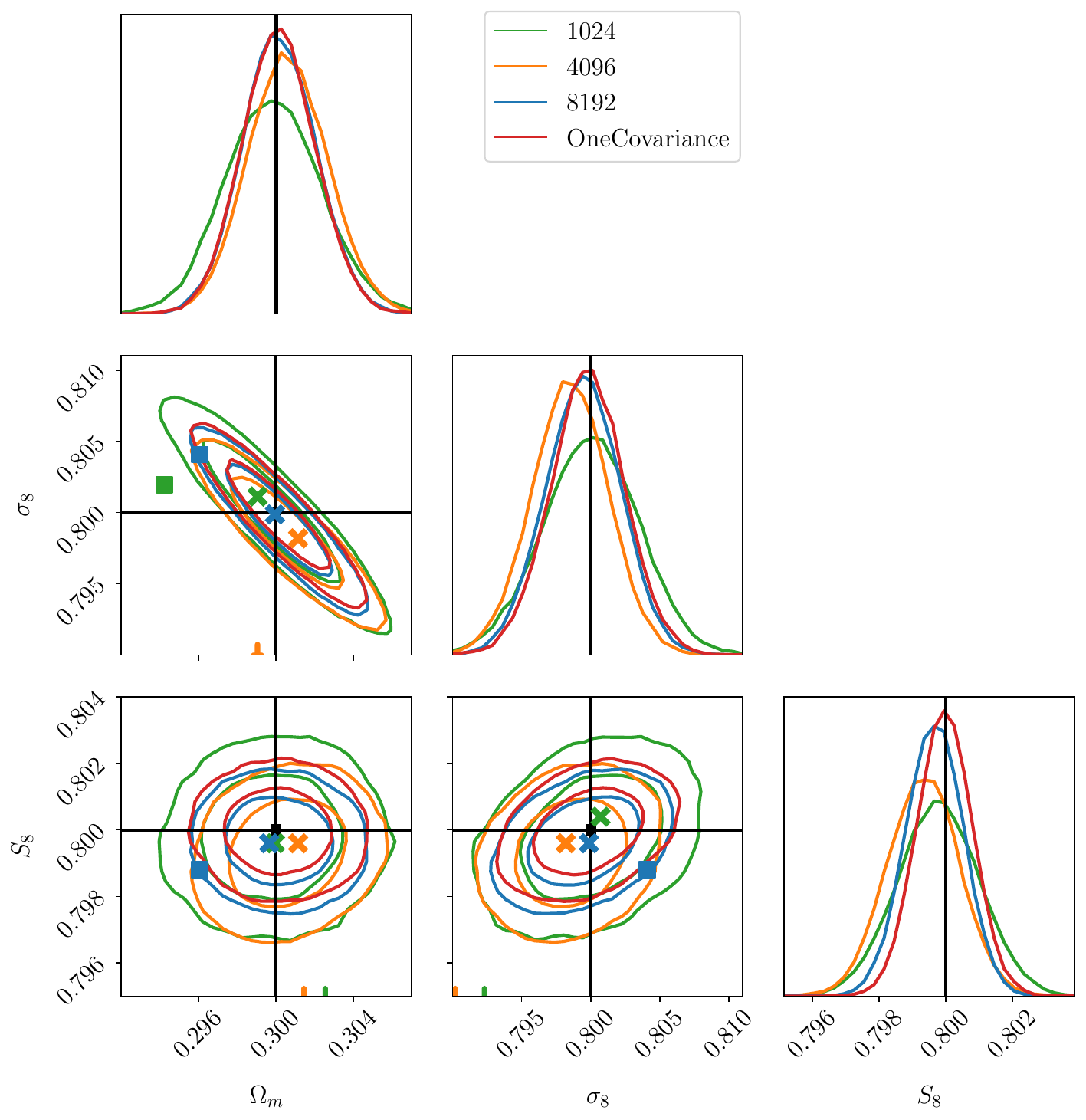}
    \caption{The constraints on parameters $\Omega_m$ and $\sigma_8$ from different  covariance matrices. The contours in different colors show the constraints from the covariance matrices predicted by $\nside = 1024$ (green), $\nside = 4096$ (orange), $\nside=8192$ (blue), and \onecovariance{} (red). The contours from the inside out represent the 1-$\sigma$ and 2-$\sigma$ confidence intervals. The black square marks the fiducial parameters $\Omega_m = 0.3$ and $\sigma_8 = 0.8$. The cross-shaped markers indicate the optimal fitting values for the corresponding simulations. The square-shaped markers show the optimal fitting values for the simulations without discarding any scales.}
    \label{fig:mcmc}
\end{figure}

The results are shown in Figure \ref{fig:mcmc}. The contours represent the 1-$\sigma$ and 2-$\sigma$ intervals for $\nside = 1024$ (green), $\nside = 4096$ (orange), $\nside=8192$ (blue), and \onecovariance{} (red). The cross-shaped markers indicate the optimal fitting values for the corresponding simulations. The square-shaped markers show the optimal fitting values for the simulations without discarding any scales.

As shown in Figure \ref{fig:mcmc}, through discarding the scales within the critical scales, the generated mocks have a high fidelity, in which the statistics from the mocks indeed represent the fiducial cosmology. The contours for $\nside = 8192$, and the ones for \onecovariance{} exhibit high consistency, which indicates that we have successfully reproduced the accurate covariance matrix from the mock-generated galaxy catalogue. Meanwhile, the resolution parameter $\nside$ still plays a crucial role. The optimal fitting value for $\nside = 8192$ is closer to the fiducial cosmology than those for $\nside = 4096$ and $1024$. When $\nside = 8192$, the estimated values of $\Omega_{m} \approx 0.30025$ and $\sigma_8 \approx 0.79945$ have a fractional error much less than 1\textperthousand \ compared to the fiducial cosmology. Figure \ref{fig:mcmc} also demonstrates the importance of discarding scales below the critical scale. Otherwise, even with the maximum resolution parameter $\nside=8192$, there would still be a relative error of about 2\%.

\section{Conclusion and Discussion}
\label{sec:conclusion}

In this work, we have explored the mock generation pipeline for the \tttpt{} analysis of the CSST using \glass{}. Our primary goals were to validate the accuracy of the \tttpt{} statistics in both spherical harmonic space and real space. 

Our validation results showed that the angular power spectrum of the overdensity fields exhibited excellent consistency with the theoretical inputs from \ccl{}, even at small scales close to the band limit. We also found that the angular power spectrum of the convergence fields maintained good agreement with the theoretical values incorporating the step-shaped lensing kernel. The step-shaped lensing kernel primarily affected the values of the angular power spectrum at large scales for low-redshift shells, and this difference gradually diminished as the number of foreground matter shells increased.

In real space, the \tttpt{} correlation functions agreed well with the theoretical values at large scales. At small scales ($\theta < 5 \text{ arcmin}$), both the count-shear correlation function $\gamma_t(\theta)$ and the shear-shear correlation function $\xi^-$ were found to be about 10\% lower than the theoretical values. This discrepancy was likely caused by numerical differences due to the finite spherical resolution. Additionally, the absolute values of $\gamma_t$ and $\xi^-$ were closer to zero at small scales, which further amplified the differences.

Finally, we employed MCMC simulations to constrain the posterior distributions of the cosmological parameters $\Omega_m$ and $\sigma_8$ using the constructed covariance matrices. We found that simulations where small scales within the critical scales had been excluded provided reliable constraints that are consistent with the fiducial cosmology. For $\nside = 8192$, the best fitting reproduced from the mocks demonstrated a fractional error less than 1\textperthousand. In contrast, simulations which included these small scales showed a significant dispersion (about 2\%) since we have to discard too much information at small scales, highlighting the importance of discarding appropriate scales.

In conclusion, this work successfully constructed a pipeline to generate high fidelity mocks, which can be used to calculate the covariance matrix for weak gravitational lensing \tttpt{} based on real-space two-point correlation functions. 
The clean mocks can accurately recover the posterior probability distributions of the cosmological parameters. We emphasized the improvement in the accuracy of the covariance matrix brought by increasing the spherical resolution. 
In the future, we will be able to introduce various systematic effects directly into the samples, allowing for a more intuitive observation of their impacts on the covariance matrix and cosmological parameter constraints.

\acknowledgments
This work was supported by the National Key R\&D Program of China  (No. 2023YFA1607800, 2023YFA1607802), the National Science Foundation of China (Grant Nos. 12273020), the China Manned Space Project with No. CMS-CSST-2021-A03 and CMS-CSST-2025-A04, the “111” Project of the Ministry of Education under grant No. B20019, and the sponsorship from Yangyang Development Fund.
This work made use of the Gravity Supercomputer at the Department of Astronomy, Shanghai Jiao Tong University.

\newpage


\bibliographystyle{JHEP}
\bibliography{Cosmology.bib}

\end{document}